\documentclass[12pt]{article}
\usepackage{amssymb,amsmath,cite,bm}
\usepackage[dvips]{graphicx,color}
\usepackage{psfrag,subfigure}
\newtheorem{remark}{Remark}

\begin{document}

% ----------------------------------------------------------------
\title{\bf Analytical Mechanics With Quasi-Velocities}

\author{Farhad Aghili\thanks{email: faghili@encs.concordia.ca}}

\date{}

\maketitle

\begin{abstract}
This paper presents a formulation of Lagrangian dynamics of constrained mechanical systems in terms of reduced quasi-velocities and quasi-forces that can be used for simulation, analysis, and control purposes. In this formulation, Cholesky decomposition of the mass matrix in conjunction with adequate orthogonal matrices are used to define reduced-quasi velocities, input quasi-forces, and constraint quasi-forces which possess natural metric. The new state and input variables always have homogeneous units despite the generalized coordinates may involve in both translational and rotational components and the constraint wrench may involve in both force and moment components. Therefore, this formulation is inherently invariant with respect to changes in dimensional units without requiring weighting matrices.  Moreover, in this formulation the equations of motion are completely decoupled from those of constrained force. This allows the possibility of a simple force control action that is totally independent of the motion control action facilitating a hybrid force/motion control. The properties of the new dynamics formulation are investigated and subsequently, force/motion tracking control and regulation of constrained multibody systems based on quasi-velocities and quasi-forces are presented.
\end{abstract}

\section{Introduction}

A unifying idea for most modelling techniques used for multibody system (MBS) dynamics
is to describe the equations of motion in terms of generalized
coordinates and generalized velocities. In classical mechanics of
constrained systems, a generalized velocity is taken to be an
element of tangential space of configuration manifold, and a
generalized force is taken to be the cotangent space. However,
neither space possesses a natural metric as the generalized
coordinates or the constrains may have a combination of rotational
and translational components. As a result, the corresponding dynamic
formulation is not invariant and a solution depends on measure units
or a weighting matrix selected
\cite{Lipkin-Duffy-1988,Manes-1992,,Aghili-Piedboeuf-2003a,Angeles-2003,Aghili-2005,Luca-Manes-1994,Aghili-2015b,Aghili-2020a}. 

Transformation of the Lagrange dynamics into quasi-Lagrange dynamics, and with feedback force/position control,  have been established in the literature \cite{McClamroch-Wang-1988,Brogliato-Niculescu-1997,Brogliato-2014}. Mathematical models for constrained robot dynamics, incorporating the effects of constraint force required to maintain satisfaction of the constraints, and tracking using feedback control were presented  in \cite{McClamroch-Wang-1988}. The problem of the control of a class of mechanical systems with a finite number of degrees-of-freedom, subject to unilateral constraints on the position is presented in \cite{Brogliato-Niculescu-1997}. Various switching control strategies are analysed in this work based on a nonsmooth dynamics formulation.  The use of kinetic quasi-velocities was also developed for a particular case in \cite{Brogliato-1996,Brogliato-2016}.  Alternatively, the dynamics of a multibody system can be formulated in terms of the vector of quasi-velocities, i.e., a vector whose Euclidean norm is proportional to the {\em square root} of the system's
kinetic energy. It is known that this formulation can lead to simplification of the equations of unconstrained MBSs
\cite{Koditschek-1985,Gu-Loh-1987,Spong-1992,Rodriguez-Delgado-1992,Aghili-2020a,Bahar-1994,Jain-Rodriguez-1995,Aghili-2021a,Kozlowski-1998,Papastavridis-1998,Gu-2000,Herman-2005,Herman-Kolowski-2006,Aghili-Buehler-Hollerbach-1997a,Lduha-Ravani-1995,Junkins-Schaub-1997,Aghili-Buehler-Hollerbach-2001,Sinclair-Hurtado-Junkins-2006,Aghili-2007c,Bedrossian-1992,Aghili-2021a}.  %Aghili-2008f
In short, the square-root
factorization of the mass matrix is used as a transformation to obtain
the quasi-velocities, which are a linear combination of the
velocity and the generalized coordinates
\cite{Papastavridis-1998,Herman-Kolowski-2006,Pila-2020,Aghili-2021a}.

In the literature, the concept of quasi-velocities has been used for dynamics modelling of unconstrained MBSs. The differential variational principle of Jourdain was extended in \cite{Bahar-1994} to cover the dynamics of impulsive motion formulated in terms of quasi-velocities. It was shown by Kodistchek \cite{Koditschek-1985} that if the
square-root factorization of the inertia matrix is integrable, then
the dynamics can be significantly simplified. In such a case,
transforming the generalized coordinates to quasi-coordinates by
making use of the integrable factorization modifies the dynamics to the system of a double integrator. It was later realized by Gu {\em et al.} \cite{Gu-Loh-1987} that
such a transformation is a canonical transformation because it
satisfies Hamilton's equations. Rather than deriving the mass matrix
of MBS first and then obtaining its factorization, Rodriguez {\em et
al.} \cite{Rodriguez-Delgado-1992} derived the closed-form
expressions of the mass matrix factorization of a MBS and its
inverse directly from the link geometric and inertial parameters.
This eliminates the need for the matrix inversion required to
compute the forward dynamics. The interesting question of when the factorization of the inertia matrix is integrable, i.e., the factorization being the Jacobian of
some quasi-coordinates, was addressed independently in
\cite{Spong-1992} and \cite{Bedrossian-1992}.  It was shown that Riemannian manifold defined by the inertia matrix should be locally flat. The advantages of using the notion quasi-velocities for control of unconstrained manipulators have been recognized by many researchers and
various setpoint PD controllers based on the quasi-velocities
feedbacks have been proposed
\cite{Jain-Rodriguez-1995,Kozolowski-Herman-2000,Herman-Kozlowski-2001,Herman-2005}.  A closed inverse dynamic formulation by the Lagrangian approach in terms of quasi-coordinates for the general Stewart platform manipulator is presented in \cite{Chen-2003}. The quasi-Lagrange equations are derived with and without friction in \cite{Brogliato-2014}  based on  quasi-velocities computed with the kinetic metric of a Lagrangian system. The quasi-Lagrange dynamics involves  the mass matrix inversion that allows for a clear splitting between normal and tangential dynamics. Generalization of Lagrangian dynamics equations by taking into account the unilateral and bilateral contacts as well as frication  can be also found in \cite{Brogliato-2016}. In spite of different quasi-Lagrange dynamics formulations proposed in the literature, preservation of homogeneous units yet needs to be rendered in these  formulations.

A problem that often arises in motion/force control of MBSs with minimum solution to joint rate or force, is that generalized coordinate  may have a combination of rotational and
translational components that can be even compounded by having
combination of rotational and translational constraints
\cite{Doty-Melchiorri-Bonivento-1993}. This may lead to inconsistent
results unless adequate weighting matrices are used
\cite{Doty-Melchiorri-Bonivento-1993,Manes-1992,Featherstone-Fijany-1999,Featherstone-Thiebaut-Khatib-1999,Aghili-2005}.
For example, the minimum joint rate rates
or minimum norm force are not meaningful
quantities if the MBS has both revolute and prismatic joints
\cite{Doty-Melchiorri-Bonivento-1993}. The contribution of this paper is to extend the concept of square-root factorization of inertia matrix to define homogeneous  vectors of quasi-velocities and quasi-forces
for dynamics formulation of constrained MBS that can be
used for simulation, analysis, and control purposes. In this paper, we introduce new state and input variables comprising of reduced quasi-velocities, input quasi-forces, and constraint quasi-forces by making use of Cholesky decomposition and adequate orthogonal matrices in order to derive Lagrangian dynamics of constrained mechanical systems.
The advantages of the square--root factorization based formulation of the constrained Lagrangian dynamics is that every  vectors of quasi-velocities,  input quasi-forces, or constraint quasi-forces all have the same physical units. Therefore, unlike other approaches
\cite{Doty-Melchiorri-1993,Luca-Manes-1994,Schutter-Bruyinckx-1996,Aghili-2005},
this formulation does not require any weighting matrix when the generalized coordinates or the constraints have both translational and rotational components.
Furthermore, the equations of motions and
the equation of constraints are decoupled in such a way that separate control
inputs are associated to each set of equations, which facilitates
motion/force control of constrained systems such as robotic
manipulators. This paper is organized as follows: Section~\ref{sec:Modeling} presents the derivation of constrained Lagrangian dynamics formulation based on the notion of reduced quasi-velocities and decoupling quasi-forces. Properties of the dynamic formulation that could be useful for control purposes are presented in Sections~\ref{sec:properties}. Finally, Section~\ref{sec:control} is devoted to force/motion control based on reduced quasi-velocities and quasi-forces.

\section{Quasi-Variables Transformation} \label{sec:Modeling}

The kinetic energy of a MBS has the following quadratic form:
\begin{equation} \label{eq:K}
K(\bm q, \dot{\bm q})=\frac{1}{2} \dot {\bm q}^T \bm M(\bm q)
\dot{\bm q},
\end{equation}
where vectors $\bm q\in \mathbb{R}^n$ and  $\dot{\bm q}\in \mathbb{R}^n$ are the  generalized
coordinates and generalized velocities, and $\bm M(\bm q)$ is the generalized inertia matrix, which is {\em
symmetric} and {\em positive definite} for all $\bm q$.
According to the {\em Cholesky decomposition}, the symmetric and
positive-definite matrix $\bm M$ can be decomposed into
\begin{equation} \label{eq:M_decompose}
\bm M = \bm Q \bm Q^T,
\end{equation}
where $\bm Q$ is a lower--triangular matrix
with strictly positive-diagonal elements; $\bm Q$ is also called
the {\em Cholesky triangle}. The following formula can be used
to obtain the Cholesky triangle through some elementary operations
\begin{align} \label{eq:Cholesky}
Q_{ii} & = \big( M_{ii} - \sum_{k=1}^{i-1} Q_{ik}^2 \big)^{1/2}
\quad  \quad \quad \forall i=1,\cdots,n \\ \nonumber Q_{ji} &= \big(
M_{ji} - \sum_{k=1}^{i-1} Q_{jk}Q_{ik} \big)/Q_{ii} \quad \quad
\forall j=i+1,\cdots,n
\end{align}
Since $\bm Q$ is a lower-triangular matrix, its inverse can be
simply computed by the back substitution technique.

The dynamics equations of a constrained MBS with kinetic energy $K$ can be derived by the {\em Euler--Lagrange} equations
\begin{subequations}
\begin{align} \label{eq:constraint_lagrange}
\frac{\rm d}{{\rm d}t} \left( \frac{\partial K}{\partial \dot{\bm
q}} \right) &- \frac{\partial K}{\partial \bm q} = \bm\tau - \bm A^T \bm\lambda\\ \label{eq:Adotq}
\bm A(\bm q) \dot{\bm q} &= \bm 0
\end{align}
\end{subequations}
Here, $\bm\tau=\bm\tau_a + \bm\tau_f +\bm\tau_p$  is the generalized forces containing all applied loads including the actuator forces applied to the joints $\bm\tau_a$ and the joint friction $\tau_f$ plus the conservative forces $\bm\tau_p=-\partial P/\partial \bm
q$ owing to gravitational energy, vector $\bm\lambda \in \mathbb{R}^m$ represents the {\em generalized Lagrangian multipliers}, and matrix $\bm A \in
\mathbb{R}^{m \times n}$ is the corresponding constraint matrix associated wit the constraints  imposed on the generalized velocities. Equations \eqref{eq:Adotq} can be representation of holonomic or non-holonomic constraints. Also, it should be pointed out that $\bm A$  is not necessarily a full-rank matrix because of the possible redundant constraints. Substituting \eqref{eq:M_decompose} into \eqref{eq:K} and
then applying \eqref{eq:constraint_lagrange} yields
\begin{align} \nonumber
\bm\tau - \bm A^T \bm\lambda &= \frac{\rm d}{{\rm d}t}\big( \bm Q \bm Q^T \dot{\bm q} \big)
- \frac{1}{2} \Big( \frac{\partial}{\partial {\bm q}} \| \bm Q^T(\bm
q) \dot{\bm q} \|^2 \Big)^T\\ \notag
&= \bm Q \frac{\rm d}{{\rm d}t} \big(\bm Q^T \dot{\bm q} \big) + \dot {\bm Q} \bm Q^T \dot{\bm q} -  \frac{\partial \big(\bm Q^T \dot{\bm q} \big) }{\partial \bm q} \bm Q^T \dot{\bm q} \\ \label{eq:LLddotq}
&= \bm Q \frac{\rm d}{{\rm d}t} \big(\bm Q^T \dot{\bm q} \big)+ \Big( \dot{\bm
Q} - \frac{\partial \big( \bm Q^T \dot{\bm q}
\big)}{\partial \bm q} \Big) \bm Q^T \dot{\bm q}
\end{align}
Define the vectors of {\em quasi-velocities} and
{\em input quasi-forces} as follows:
\begin{subequations} \label{eq:quasi}
\begin{align} \label{eq:v_def}
\bm v & \triangleq \bm Q^T(\bm q) \dot{\bm q} \\ \label{eq:u_def}
\bm u & \triangleq \bm Q^{-1}(\bm q) \bm\tau,
\end{align}
\end{subequations}
\begin{remark}
Since $\det{\bm Q}=\sqrt{\det{\bm M}}\neq 0$, matrix $\bm Q$ is always full rank and thus $\bm Q^{-1}$ always exists. Therefore, \eqref{eq:quasi} implies that there are  one-to-one  relationships between the set of generalized velocity and generalized force $\{ \dot{\bm q}, \bm\tau \}$ on one hand and the set of  quasi-velocities  and quasi-forces $\{ \bm v, \bm u\}$ on the other hand.
\end{remark}
Now, pre-multiplying both sides of \eqref{eq:LLddotq} by
$\bm Q^{-1}$ and then substituting \eqref{eq:quasi} into the
resultant equation, we arrive at the equations of mechanical systems
expressed by the quasi-variables:
\begin{subequations} \label{eq:quasi_dyn}
\begin{align} \label{eq:dv=f}
\dot {\bm v} + \bm\Gamma \bm v & = \bm u - \bm\Lambda^T \bm\lambda \\ \label{eq:Qv=0}
\bm\Lambda \bm v &= \bm 0
\end{align}
\end{subequations}
where
\begin{align} \label{eq:Omega}
\bm\Gamma \triangleq & \bm Q^{-1} \Big( \dot{\bm Q} - \frac{\partial
\bm v}{\partial \bm q} \Big) \\ \label{eq:Lambda}
\bm\Lambda  \triangleq & \bm A \bm Q^{-T}
\end{align}
Alternatively, matrix $\bm\Gamma$ can be  described by $\bm\Gamma = \bm Q^{-1}\bm\Psi$
where the $ij$th entries of matrix $\bm\Psi$ can be calculated through the following partial derivative equations
\begin{equation} \label{eq:Qij}
\Psi_{ij} = \sum_k \Big( \frac{\partial  Q_{ij}}{\partial q_k} - \frac{\partial Q_{kj}}{\partial q_i} \Big) \dot q_k.
\end{equation}
It is worth noting that the constraint equations \eqref{eq:Qv=0} are imposed on the quasi-velocities analogous to constraint equations \eqref{eq:Adotq} imposed on the generalized velocities. Since matrix $\bm Q$ is always full-rank, we can say
$\mbox{rank}(\bm\Lambda)=\mbox{rank}(\bm A)=r$, where $r\leq m$ is
the number of independent constraints. Then, according to the {\em
singular value decomposition} (SVD) there exist unitary (orthogonal)
matrices $\bm U=[\bm U_1 \;\; \bm U_2]\in \mathbb{R}^{m \times m}$
and $\bm V=[\bm V_1 \;\; \bm V_2]\in \mathbb{R}^{n \times n}$ (i.e.,
$\bm U^T \bm U = \bm U \bm U^T = \bm I_m$ and $\bm V^T \bm V = \bm V \bm V^T = \bm I_n$) such that
\begin{equation} \label{eq:svd}
\bm\Lambda = \bm U \bm\Sigma \bm V^T \quad \text{where} \quad
\bm\Sigma =
\begin{bmatrix} \bm S & \bm 0 \\ \bm 0 & \bm 0 \end{bmatrix}
\end{equation}
and $\bm S= \mbox{diag}(\sigma_1 , \cdots, \sigma_r)$ with $\sigma_1
\geq \cdots \geq \sigma_r > 0$ being the non-zero singular values
\cite{Klema-Laub-1980,Press-Flannery-1988}. The unitary matrices are
partitioned so that the dimensions of the submatrices $\bm U_1$ and
$\bm V_1 \in \mathbb{R}^{n \times r}$ are consistent with those of $\bm S$. That is the columns
of $\bm U_1$ and $\bm V_2 \in \mathbb{R}^{n \times (n-r)}$ are the corresponding sets of orthonormal
eigenvalues which span the range space and the null space of
$\bm\Lambda$, respectively \cite{Golub-VanLoan-1996}. Thus
\begin{equation}  \label{eq:LambdaV2}
\bm\Lambda \bm V_2^T = \bm 0
\end{equation}
Define {\em reduced order quasi-velocities} or {\em independent quasi-velocities}  $\bm v_r \in \mathbb{R}^{n-r}$ as follow:
\begin{align} \notag
\bm v_r &= \bm V_2^T \bm v \\ \label{eq:vr=V2v}
&= \bm V_2^T \bm Q^T \dot{\bm q}
\end{align}
Since $\bm V_2 \bm V_2^T$ is a projection matrix corresponding to the kernel of matrix $\bm\Lambda$, we have  $\bm V_2 \bm V_2^T\bm v = \bm v$. Therefore, we can readily obtain the reciprocal of \eqref{eq:vr=V2v} by per-multiplying both sides of the latter equation by $\bm V_2$, i.e.,
\begin{equation} \label{eq:v=V2vr}
\bm v= \bm V_2 \bm v_r
\end{equation}
The time-derivative of the reduced quasi-velocities can be expressed by
\begin{equation} \label{eq:dot_vr}
\dot{\bm v}_r = \bm V_2^T \dot{\bm v} + \dot{\bm V}_2^T \bm V_2 \bm v_r
\end{equation}
It can be inferred from  \eqref{eq:v_def} and \eqref{eq:v=V2vr} that the
generalized velocities can be constructed from the reduced quasi-velocities via the following mapping
\begin{equation} \label{eq:dotq=Qvr}
\dot{\bm q} = \bm Q^{-T} \bm V_2 \bm v_r
\end{equation}
Finally, pre-multiplying both sides of \eqref{eq:dv=f} by $\bm V_2^T$ and then using identities \eqref{eq:LambdaV2},  \eqref{eq:v=V2vr}, and \eqref{eq:dot_vr}, we arrive at
\begin{equation} \label{eq:dot_v_r}
\dot {\bm v}_r = \bm\Gamma_v \bm v_r + \bm u_v
\end{equation}
where
\begin{subequations}
\begin{align} \label{eq:Gamma_v}
\bm\Gamma_v &= \bm V_2^T \bm\Gamma \bm V_2 + \dot{\bm V}_2^T \bm V_2 \\ \label{eq:u_v}
\bm u_v &= \bm V_2^T \bm Q^{-1} \bm\tau
\end{align}
\end{subequations}
Note that matrix $\dot{\bm V}_2^T$ in the RHS of \eqref{eq:Gamma_v} can be obtained from time-derivative of \eqref{eq:LambdaV2} as follows: $\bm\Lambda \dot{\bm V}_2^T = - \dot{\bm\Lambda} \bm V_2^T$, and thus
\begin{equation} \notag
\dot{\bm V}_2^T = - \bm V \bm\Sigma^{+} \bm U^T \dot{\bm\Lambda}\bm V_2^T
\end{equation}
where $\bm\Sigma^{+}$ contains the inverse of non-zero singular values. On the other hand, using the orthogonality property $\bm V_1^T \bm V_2 = \bm0$ and pre-multiplying both sides of \eqref{eq:v=V2vr} by $\bm V_1^T$, one can conclude that $\bm V_1^T$ indeed acts an annihilator for the quasi-velocities, i.e., $\bm V_1^T \bm v = \bm 0$. Thus, time-derivative of the latter identity gives us
\begin{equation} \label{eq:V1dotv}
\bm V_1^T \dot{\bm v} + \dot{\bm V}_1^T \bm v = \bm0
\end{equation}
Pre-multiplying both sides of \eqref{eq:dv=f} by $\bm V_1^T$ and then using identity \eqref{eq:V1dotv}, we arrive at
\begin{equation} \label{eq:xi}
\bm\xi  = \bm\Gamma_{\xi} \bm v_r + \bm u_{\xi}
\end{equation}
where
\begin{subequations}
\begin{equation} \label{eq:Gamma_xi}
\bm\Gamma_{\xi} = - \bm V_1^T \bm\Gamma \bm V_2 + \dot{\bm V}_1^T \bm V_2 ,
\end{equation}
and $\bm\xi $  and $\bm u_{\xi}$ are, respectively, the constraint quasi-forces and the corresponding input quasi-forces defined by
\begin{align}
\bm\xi &=  \bm V_1^T \bm Q^{-1} \bm A^T \bm\lambda\\ \label{eq:u_xi}
\bm u_{\xi} &= \bm V_1^T \bm Q^{-1} \bm\tau
\end{align}
\end{subequations}
Equations \eqref{eq:dot_v_r} and \eqref{eq:xi} completely characterize the dynamics behaviour of a constrained MBS in terms of quasi variables. It should be noted that that the corresponding input quasi-forces for the equations of motion and the constraint forces, i.e., $\bm u_v$ and $\bm u_{\xi}$ are naturally decoupled. The decoupling of the equations of motions and constraint forces allows the development of independent motion and force controllers without any need to compensate for the cross-coupling terms. It should be also pointed out that the conventional transformation used in quasi-Lagrangian dynamics  allows for splitting between the normal and tangential dynamics \cite{Brogliato-2014}. However, the normal dynamics is eliminated from the above equations, which are expressed in terms of the reduced quasi-velocities.

\subsection{Properties of the System  in Terms of Quasi-Velocities and Quasi-Forces} \label{sec:properties}
In the following analysis, we explore some properties of the quasi-variable formulation \eqref{eq:dot_v_r} and \eqref{eq:xi}
that will be useful in control design purposes.
\begin{enumerate} \label{eq:kinetic_vr}
\item Kinetic energy
\begin{equation} \label{K=vr^2}
K = \frac{1}{2} \| \bm v_r \|^2.
\end{equation}
\item Skew-symmetric property
\begin{equation}  \label{eq:skew}
\bm v_r^T \bm\Gamma_v \bm v_r =0 \qquad \forall \bm v_r \in \mathbb{R}^{n-r}.
\end{equation}
\item Boundedness of matrices $\bm\Gamma_v$ and $\bm\Gamma_{\xi}$, i.e.,
\begin{equation} \label{eq:boundedGam}
\|\bm\Gamma_v \|, \; \|\bm\Gamma_{\xi} \| \leq \gamma \| \bm v_r \|  \qquad \exists \gamma>0.
\end{equation}
\end{enumerate}
The kinetic energy  is trivially given by
\begin{equation} \label{eq:T_normv}
K = \frac{1}{2} \| \bm v \|^2 = \frac{1}{2} \bm v_r \bm V_2 ^T \bm V_2 \bm v_r = \frac{1}{2} \| \bm v_r \|^2,
\end{equation}
and hence \eqref{K=vr^2} is proven. Furthermore, in the absence of any external
active force, the principle of conservation of kinetic energy dictates that
the kinetic energy of mechanical system is bound to be constant,
i.e., $\bm u_v= \bm 0 \Longrightarrow \dot K =0$. On the other hand,
the zero-input response of a mechanical system is $\dot {\bm v} = -
\bm\Gamma_v \bm v_r$. Substituting the latter equation in the
time-derivative of \eqref{eq:T_normv} gives $\bm v_r^T \bm\Gamma_r \bm v_r =  0$, which implies \eqref{eq:skew}.
Assume that $c_m$ denote the minimum eigenvalue of $\bm M$ for all
configurations $\bm q$, that is, $c_m \bm I \leq \bm M(\bm q)$.
Then, using the norm properties leads to
\begin{equation} \label{eq:norm_invW}
\| \bm Q^{-1} \| \leq \frac{1}{\sqrt{c_m}}.
\end{equation}
In view of \eqref{eq:dotq=Qvr}  and \eqref{eq:norm_invW} and knowing
that $\| \bm V_2 \|=1$, we can say
\begin{equation} \label{eq:bound_dq}
\| \dot{\bm q} \| \leq \| \bm Q^{-T} \| \| \bm V_2 \| \| \bm v_r \|
\leq \frac{\|\bm v_r \|}{\sqrt{c_m}}.
\end{equation}
Moreover, if the factorization $\bm Q(\bm q)$ is a sufficiently
smooth function, then all partial-derivative terms in \eqref{eq:Qij} are bounded
and hence there exists a finite $c_{\psi}> 0$ such that $\| \bm\Psi \| \leq
c_{\psi} \| \dot{\bm q} \|$, and hence we can say
\begin{equation} \label{eq:norm_Gamma}
\| \bm\Gamma \| \leq \frac{c_{\psi}}{c_m} \| \bm v_r \|.
\end{equation}
Also, the entries of the time-derivative of matrix $\bm V$ can be written as
\begin{equation}
\dot{V}_{ij} = \sum_k \frac{\partial V_{ij} }{\partial q_k} \dot q_k
\end{equation}
Consequently, we can say $\| \dot {\bm V} \| \leq c_V \| \dot{\bm q} \| \leq c_V/c_m \| \bm v_r \|$. Finally, knowing that  $\|\bm V_1 \| = \| \bm V_2 \|=1$ and $\| \dot{\bm V}_1 \| , \| \dot{\bm V}_2 \| \leq \| \dot{\bm V} \|$, one can infer \eqref{eq:boundedGam} from the RHS expressions in \eqref{eq:Gamma_v} and \eqref{eq:Gamma_xi} where $\gamma =(c_V + c_{\psi})/c_m$.

\subsection{Natural Metric}
A problem that often arises in robotics, namely hybrid control or
the minimum solution to joint rate or force is that generalized
coordinate $\bm q$ may have a combination of rotational and
translational components that can be even compounded by having
combination of rotational and translational constraints
\cite{Doty-Melchiorri-Bonivento-1993}. This may lead to inconsistent
results, i.e., results that are invariant with respect to changes in
dimensional units unless adequate weighting matrices are used
\cite{Doty-Melchiorri-Bonivento-1993,Manes-1992,Featherstone-Fijany-1999,Featherstone-Thiebaut-Khatib-1999,Aghili-2005}.
For example, the minimum joint rate rates, $\min \| \dot{\bm q} \|$,
or minimum norm force, $\min \| \bm\tau \|$, are not meaningful
quantities if the robot has both revolute and prismatic joints
\cite{Doty-Melchiorri-Bonivento-1993}.
\begin{remark}
It is also important to note the important property of the reduced quasi-velocities and quasi-forces is that they always have homogeneous units.
The expression of kinetic energy \eqref{K=vr^2} implies  $ \| \bm v_r \| = \| \bm v \| = \sqrt{2K}$ meaning that  all elements of the vector of quasi-velocities $\bm
v$ or $\bm v_r$ must have a homogeneous unit $[\sqrt{\rm kg}{\rm
m}/{\rm s}]$. This is true even if the vector of the generalized
coordinate or the constraints have combinations of rotational and
translational components. Similarly, one can argue from \eqref{eq:dot_v_r} and \eqref{eq:xi}
 that the elements
of the quasi-forces $\bm u_v$ and $\bm\xi$ have always identical unit $[\sqrt{\rm kg}{\rm
m}/{\rm s}^2]$, regardless of the units of the generalized force or
the constraint wrench.
\end{remark}

Therefore, minimization of  $\| {\bm v} \|$
or $\min \| \bm u \|$ is legitimate because the latter vectors have
always homogeneous units. Moreover, the selection matrices which are
often needed in hybrid position-force control of manipulators when
both translational and rotational constraints are involved between
its end effector and its environment~\cite{Featherstone-Thiebaut-Khatib-1999} becomes a non-issue here.

\subsection{Existence of Quasi-Coordinates} \label{sec:quasi-coordinates}

It should be pointed out that despite the one-to-one
correspondence between velocity coordinate $\dot {\bm q}$ and the
quasi-velocities $\bm v$, they are not synonymous. This is because
the integration of the former variable leads to the generalized
coordinate, while integration of the latter variable does not always lead
to a meaningful vector describing the configuration of the
mechanical system. Now let us assume that $\dot{\bm\phi} = \bm v$, where ${\bm\phi}$ is called {\em
quasi-coordinates}. For $\bm\phi$ to be an explicit function of $\bm
q$, i.e., $\bm\phi = \bm\phi(\bm q)$, it must be the gradient of a
scalar function meaning that $\bm\phi$ is a {\em conservative vector
field}. In that case, \eqref{eq:v_def} implies that $\bm Q^T(\bm q)$
is actually a Jacobian as
\begin{equation}
Q_{ij}= \frac{\partial \phi_j}{\partial q_i}.
\end{equation}
If $\bm\phi(\bm q)$
exists and it is a smooth function, then we can say
\begin{equation}
\frac{\partial Q_{ij}}{\partial q_k} - \frac{\partial Q_{kj}}{\partial q_i}= \frac{\partial^2 \phi_j}{\partial q_i \partial q_k} - \frac{\partial^2 \phi_i}{\partial q_k \partial q_i} =0
\end{equation}
Under this circumstance, the expression in the
parenthesis of the right-hand side of \eqref{eq:Qij} vanishes, i.e., $\bm\Psi \equiv \bm 0$ and hence $\bm\Gamma \equiv \bm 0$. Therefore, the equations of motion become a simple
integrator system and hence $\bm\phi$ and $\bm v$ are
indeed alternative possibilities for generalized coordinates and
generalized velocities. Technically speaking, a necessary and sufficient condition for the
existence of the quasi-coordinates, $\bm\phi$, is that the
Riemannian manifold defined by the inertia matrix $\bm M(\bm
q)$ be locally flat--by definition, a Riemannian manifold
that is locally isometric to Euclidean manifold is called a locally
flat manifold \cite{Spong-1992}. However, that has been proven to
be a very stringent condition \cite{Bedrossian-1992}. Nevertheless, state variables $\{
\bm q,  \bm v_r \}$ are sufficient to describe
completely the states of MBS. Setting \eqref{eq:dotq=Qvr} and \eqref{eq:dot_v_r}  in state space
form gives
\begin{equation} \label{eq:ode}
\frac{\rm d}{{\rm d}t} \begin{bmatrix} \bm q \\ \bm v_r \end{bmatrix}
=
\begin{bmatrix}  \bm Q^{-T}\bm V_2
\\ - \bm\Gamma_v \end{bmatrix} \bm v_r  +
\begin{bmatrix} \bm 0 \\ \bm I \end{bmatrix} \bm u_r.
\end{equation}
It is interesting to note that dynamics system \eqref{eq:ode} is in
the form of the so-called {\em second-order kinematic model} of
constrained mechanism, which appears in kinematics of nonholonomic
systems. This is the manifestation of the fact that the integration
of quasi-velocities, in general, does not lead to
quasi-coordinates.

\section{Force/Motion Control Based on Quasi-Velocities and Quasi-Forces} \label{sec:control}

In this section, we use dynamics formulation  \eqref{eq:dot_v_r} and \eqref{eq:xi} for tracking control and regulation control of constrained MBSs. Due to presence of only $r$ independent constraints, the actual number of degrees of freedom of the system is reduced to $n-r$.
Consequently, there must be $n-r$ independent variables
$\bm\theta(\bm q)\in \mathbb{R}^{n-r}$, which is also called a {\em
minimal set of generalized coordinates}. Thus, we can say
\begin{equation} \label{eq:dot_tet}
\dot{\bm\theta} = \left( \frac{\partial \bm\theta}{\partial \bm q}  \right) \dot{\bm q}
\end{equation}
Substituting $\dot{\bm q}$ from \eqref{eq:dotq=Qvr} into \eqref{eq:dot_tet} yields
\begin{equation} \label{eq:dtheta}
\dot{\bm\theta}  =  \bm D(\bm\theta) \bm v_r, \qquad \text{where}
\qquad \bm D \triangleq \left( \frac{\partial \bm\theta}{\partial \bm q} \right)
\bm Q^{-T} \bm V_2.
\end{equation}
Since  both variables $\bm v_r$ and $\dot{\bm\theta}$ are
with the same dimension, the reciprocal of mapping \eqref{eq:dtheta}
must uniquely exist, i.e.,  $ \bm D^{-1}$ is always well-defined. Now we adopt a Lyapunov-based  control scheme \cite[p.74]{Canudas-Siciliano-Bastin-book-1996} for designing a feedback
control in terms of quasi-velocities. Define the composite error
\begin{equation} \label{eq:edef}
\bm\epsilon \triangleq
\bm\epsilon_{v_r} + k_p \bm D^{-1} \bm\epsilon_{\theta} ,
\end{equation}
where $k_p>0$, $\bm\epsilon_{v_r}={\bm v}_r - {\bm v}_{r_d}$, and
$\bm\epsilon_{\theta} =\bm\theta - \bm\theta_d $. Also, define the auxiliary
variable $\bm\sigma = \bm v_r -\bm\epsilon =   \bm v_{r_d} - k_p\bm D^{-1}
\bm\epsilon_{\theta}$, which is used in the following control law:
\begin{equation} \label{eq:contrlaw_zeta}
\bm u_v = \dot{\bm\sigma} + \bm\Gamma_v \bm v_r - k_d \bm\epsilon,
\end{equation}
where $k_d >0$. Applying control law \eqref{eq:contrlaw_zeta} to
system \eqref{eq:quasi_dyn} gives the dynamics of the error $\epsilon$
in terms of the following first-order differential equation:
\begin{equation} \label{eq:diff_eps}
\dot{\bm \epsilon} = - k_d \bm\epsilon.
\end{equation}
In other words,  the composite error $\bm\epsilon$ is exponentially stable
\begin{equation} \label{eq:eps_norm}
\bm\epsilon  =  \bm\epsilon (0)   e^{-k_d t}.
\end{equation}
Pre-multiplying both sides of \eqref{eq:edef} by $\bm D(\bm\theta)$,
the resultant equation can be rearranged to the following
differential equation
\begin{equation} \label{eq:dot_tildeq}
\dot{\bm\epsilon}_{\theta}   =  - k_p \bm\epsilon_{\theta} +  \bm D
\bm \epsilon.
\end{equation}
Now, it remains to show that the solution of the above
non-autonomous system converges to zero. First, notice from definition  \eqref{eq:dtheta} that $\bm D$  is the product of three bounded matrix and thus it should be a bounded matrix too. That is because $\| \bm Q^{-T} \|\leq 1/ \sqrt{c_m}$ according to \eqref{eq:norm_invW} and $\bm V_2$ is a unitary matrix meaning that $\| \bm V_2 \| \leq 1$. Thus,
there exists scalar $c_d>0$ such that
\begin{equation} \label{eq:bounded_B}
\bm D(\bm\theta)  \leq c_d \bm I.
\end{equation}
where $c_d=\| \partial \bm Q/ \partial \bm q \| / \sqrt{c_m}$. One can show that the solution of \eqref{eq:dot_tildeq} satisfies
\begin{equation} \label{eq:solution_normdq}
\| \bm\epsilon_{\theta}  \| \leq \Big( \| \bm\epsilon_{\theta} (0) \|  + c_d \| \bm\epsilon(0) \| \Big) e^{-k_d t},
\end{equation}
which implies exponential stability of the tracking error; see Appendix for details.  Tracking of the desired constraint force $ \bm\xi_d = \bm\Lambda_r^T \bm\lambda_d$ can be
simply achieved by compensating the velocity perturbation term
in \eqref{eq:xi}, i.e.,
\begin{equation}
\bm u_{\xi}  = \bm\xi_d + \bm\Gamma_{\xi}  \bm v_r \quad \Longrightarrow \quad \bm\xi = \bm\xi_d.
\end{equation}
From definitions of $\bm u$, $\bm u_v$, and $\bm u_{\xi}$ in \eqref{eq:u_def}, \eqref{eq:u_v} and \eqref{eq:u_xi}, one can write the following relationship
\begin{equation} \notag
\begin{bmatrix} \bm u_{\xi} \\ \bm u_{v} \end{bmatrix} = \bm V^T \bm u
\end{equation}
and thus $\bm u_{\xi}^T \bm u_{\xi} + \bm u_v^T \bm u_v = \bm u \bm V \bm V^T \bm u$. Since $\bm V$ is a unitary matrix, the latter identity is equivalent to $\|\bm u_v \|^2 + \|\bm u_{\xi} \|^2 = \|\bm u \|^2$. It is worth noting that in view of the latter norm identity , we can say that
\begin{equation} \notag
\min \| \bm u \| \quad \leftarrow \quad \bm u_{\xi} \equiv \bm 0
\end{equation}
That is tantamount to minimization of weighted norm of the generalized forces where the
weight matrix is the inverse of the inertia matrix because
\begin{equation} \label{eq:norm_u}
\|\bm u \|= \sqrt{\bm\tau^T \bm M^{-1} \bm\tau}.
\end{equation}
In other word, the kinetic metric of the generalized force in minimized.

\subsection{Setpoint Control}

In this section, we extend such a feedback control for hybrid
motion/force control of constrained robotic systems. Consider the following control law for system \eqref{eq:quasi_dyn}
\begin{equation} \label{eq:control}
\bm u_v =  - k_d \bm v_r - k_p \bm\epsilon_{\theta} ,
\end{equation}
where $\bm\epsilon_{\theta} =\bm\theta - \bm\theta_d$, and  $k_d,k_p>0$. Then, the dynamics of the closed--loop system becomes
\begin{equation} \label{eq:closed_reps}
\dot{\bm v}_r = - \bm\Gamma_v \bm v_r - k_d \bm v_r - k_p \bm\epsilon_{\theta} .
\end{equation}
Choose the following standard Lyapunov function
\begin{equation} \label{eq:Lyap}
V= \frac{1}{2} \| \bm\epsilon_{v_r}  \|^2 + \frac{1}{2} k_p \| \bm\epsilon_{\theta} \|^2.
\end{equation}
Then, using Property~\eqref{eq:skew} in the time-derivative of
\eqref{eq:Lyap} along \eqref{eq:closed_reps} yields
\begin{equation} \notag
\dot V = - \bm v_r^T \bm K_d \bm v_r \leq 0,
\end{equation}
which is negative-semidefinite.  Therefore, according to LaSalle's Global Invariant Set
Theorem \cite{Lasalle-1960}, \cite[p.115]{Khalil-1992}, the solution
of system~\eqref{eq:closed_reps} asymptotically converges to the
invariant set ${\cal S}= \{\bm
v_r, \bm\epsilon_{\theta} : \bm v_r= \bm 0 ,  \bm\epsilon_{\theta} =\bm 0 \}$, i.e., $\bm\theta \rightarrow \bm\theta_d$ as $t$ goes to infinity. Define $\bm\xi_d=\bm\Lambda_r^T \bm\lambda_d$ where is the desired value of the Lagrangian multiplier. Then the force control law can be simply given by
\begin{equation} \label{eq:lambda_d}
\bm u_{\xi} = \bm\xi_d.
\end{equation}
Substituting \eqref{eq:lambda_d} into \eqref{eq:xi}  and using
Property~\eqref{eq:boundedGam}, we get
\[ \| \bm\xi - \bm\xi_d \| \leq \gamma \| \bm v_r \|^2. \]
Since $\bm v_r~\rightarrow~0$, then
$\bm\xi~\rightarrow~\bm\xi_d$ as $t$ goes to infinity.

\section{Conclusions}
A consistent  formulation for modelling of constrained MBSs  using
the concept of quasi-velocities and quasi-forces has been presented. The main advantage of this formulation is that it does not require adequate weighting matrices when the generalized coordinates involve in both translational and rotational components and/or the generalized force or constraint wrench involve in both force and moment components. It has been also shown that using the Cholesky decomposition of the mass matrix and the unitary transformation corresponding to the kernel of the Pfaffian constraints of the quasi-velocities led to the decoupling of motion and force control inputs. This allowed the possibility to develop a simple force control action that is totally independent of the motion control action. Some properties of the constrained Lagrangian dynamics formulation based on the quasi-variables were presented that could be useful for control purposes. It  followed by the development of the force/motion control of constrained MBSs based on the quasi-velocities and quasi-forces.

%\appendix

%--------------------------------------------------------
\section*{Appendix}
%--------------------------------------------------------
Consider the following positive--definite function
\begin{equation} \label{eq:positive_function}
V = \frac{1}{2} \| \bm\epsilon_{\theta}  \| ^2,
\end{equation}
whose time-derivative along \eqref{eq:dot_tildeq} gives
\begin{equation} \notag
\dot V = - k_p \| \bm\epsilon_{\theta}  \|^2 + \bm\epsilon_{\theta}^T \bm D(\bm\theta) \bm\epsilon.
\end{equation}
From \eqref{eq:eps_norm} and \eqref{eq:bounded_B}, one can find a
bound on $\dot V$ as
\begin{equation}
\dot V  \leq  - 2 k_p V +  c_d  \| \bm\epsilon(0) \|  \| \bm\epsilon_{\theta}  \| e^{-k_p t},
\end{equation}
which is in the form of a Bernoulli differential inequality. The
above nonlinear inequality can be linearized by the following change
of variable $U=\sqrt{V}$, i.e.,
\begin{equation} \label{eq:dU_inequality}
\dot U \leq - k_p U + \frac{c_d \| \bm\epsilon(0) \|}{\sqrt{2}}
e^{-k_p t}
\end{equation}
In view of the comparison lemma \cite[p. 222]{Khalil-1992} and
\eqref{eq:eps_norm}, one can show that the solution of
\eqref{eq:dU_inequality} must satisfy
\begin{equation} \notag
U \leq \Big( U(0) + \frac{c_d \| \bm\epsilon(0)
\|}{\sqrt{2}} \Big)  e^{- k_p t},
\end{equation}
which is equivalent to \eqref{eq:solution_normdq}.

%-------------------------------------------------------------
\bibliographystyle{IEEEtran}
%\bibliography{references}
%\end{document}

\end{document}